\documentclass[12pt]{iopart}
\usepackage{graphicx,bm}
\usepackage{amssymb,color}

\newcommand{\be}{\begin{equation}}
\newcommand{\ee}{\end{equation}}

\newcommand{\J}{{\mathbf J}}
\newcommand{\R}{{\mathbf R}}

\newcommand{\opR}{{\hat{R}}}
\newcommand{\opT}{{\hat{T}}}
\newcommand{\oprho}{{\hat{\rho}}}

\newcommand{\mH}{{\mathbf H}}

\newcommand{\M}{{\mathbf M}}

\newcommand{\GO}{{\mathcal O}}

\begin{document}

\title{Blind spots between quantum states}

\author {Eduardo Zambrano\footnote{zambrano@cbpf.br} and Alfredo M. Ozorio de Almeida\footnote{ozorio@cbpf.br}}
\address{Centro Brasileiro de Pesquisas Fisicas,
Rua Xavier Sigaud 150, 22290-180, Rio de Janeiro, R.J., Brazil}
\date{\today}

\begin{abstract}

The overlap of a large quantum state with its image, under tiny translations, oscillates swiftly.
We here show that complete orthogonality occurs generically at isolated points.
Decoherence, in the Markovian approximation, lifts the correlation minima from zero
much more quickly than the Wigner function is smoothed into a positive phase space distribution.
In the case of a superposition of coherent states, blind spots depend strongly on positions and amplitudes
of the components, but they are only weakly affected by relative phases
and the various degrees and directions of squeezing.
The blind spots for coherent state triplets are special in that they lie close to an hexagonal lattice:
Further superpositions of translated triplets, specified by nodes of one of the sublattices,
are quasi-orthogonal to the original triplet and to any state,
likewise constructed on the other sublattice.

\end{abstract}


\section{Introduction}

Interference between quantum states has no equivalence
in classical mechanics. The simplest way to generate it
is by superposing identical copies of the same initial state,
displaced in position or momentum, in eg. a generalized two slit experiment.
One produces finer fringes, the larger the separation between the states,
that is, large separation in momentum generates fine fringes
in the position intensities and vice versa.
Both these cases give rise to oscillations of the Wigner function \cite{Wigner},
which represents the superposed state in phase space: If both
copies have individual Wigner functions that are well localized in phase space,
the oscillations that lie between them are most pronounced when the individual states
are sufficiently separated for the overlap between them to be very small.
The fringes disappear as the copies are brought together.

The phenomena studied here belong to an opposite regime,
that is, though we still consider the displacement of identical copies,
their individual extent is assumed to be much larger than their separation,
i.e. their phase space areas are much larger than $\hbar$.
Rather than the intensity of probability in position or momentum, it is the
overlap itself between the pair of states, such as produced in an interferometer,
that oscilates as a function of the displacement.
Even though one generally expects large overlaps for small translations,
one may even obtain complete orthogonality with the initial state.
By increasing the extent of the state, one thus maximizes the effect
of a very small translation. We refer to displacements for which the overlap is zero as
{\it quantum blind spots}. It is as if the state were oblivious to its translation,
right beside it, `stepping on its toes'.

A displacement may be generated continuously, by an interaction
with a two-level quantum system, as in the example of Alonso \emph{et. al.} \cite{Alonso},
where the blind spots are related with the complete decoherence.
The action of a quantum Hamiltonian, approximated as a linear function in position and momentum
(in the classically small) region where the state is defined, also provides such a translation.

The oscillations of the overlap of an intial state with its evolving translations
can then be used to measure parameters of the driving Hamiltonian.
Toscano \emph{et. al.} \cite{ToscZur} have shown that the standard quantum limit (SQL)
for the precision of quantum measurements \cite{qmetrol} can then be surpassed by choosing
a specific initial state: a {\it compass state}, i.e. a particular superposition of
separated coherent states.

We here study the overlap structure of this and alternative extended states
with all their possible translations.
Is the overlap specially sensitive to decoherence?
How delicate is the balance between the locations, the amplitudes
and the phases of a superposition of coherent states for a blind spot to arise, or to disappear?
What is the effect of squeezing and rotating the component coherent states?
Do other extended states, such as excited states of
anharmonic oscillators exhibit blind spots?

It is found that blind spots resulting from the overlap of a translation are a general feature
of any extended quantum state. It is only if the state has a generalized
parity symmetry, i.e. if it is invariant with respect
to the reflection through a phase space point, that overlap zeroes may occur
along continuous displacement lines, rather than at isolated points.
The ubiquitous {\it Schr\"odinger cat} states, as well as the compass state,
are then understood to be symmetric counter examples.

Regular, or irregular patterns of blind spots manifest
delicate wavelike properties of quantum mechanics.
The rich structures that emerge
remind one of {\it quantum carpets} \cite{carpet},
though more akin to so called {\it sub-Planckian} features of pure quantum states \cite{Zurek}.
A striking consequence of purity, the invariance of the correlations
with respect to Fourier transformation \cite{ChouVou, AlmVal04, LNP}
is a prime manifestation of the conjugacy of large and small scales
and turns out to be essential to the following analysis.
In principle, the zeroes featured here may be directly observable
through manipulation and interference of simple quantum states,
such as carried out in quantum optics \cite{Leonhardt, Schleich}.
Though generically isolated, blind spots are not constrained by analyticity,
as are the zeroes of the Bargman function and, hence, the Husimi function \cite{LebVor}.
Unlike those, which often lie in shallow evanescent regions \cite{TosAlm99},
we here deal with sharp indentations on a background of maximal correlations.

In section 2 we review the general theory for the Wigner function and its Fourier transform,
the {\it characteristic function}, or the {\it chord function}.
These complete representations
of quantum mechanics are fundamentally based on reflection and translation operators,
as discovered by Grossmann \cite{Grossmann} and Royer \cite{Royer}.
They thus provide the natural setting for the present theory.
The overlap between translated pure states is then related
to a phase space correlation function, which is also meaningful for the mixed states
of open systems. In section 3,  the nodal lines of the real and
imaginary part of the chord function are analyzed: Their intersection
determines isolated blind spots, in the absence of a reflection symmetry.

A theory for the blind spots of superpositions of coherent states
or squeezed states, i.e. Schr\"odinger cat states and their generalizations,
is developed in section 4. Such superpositions of several coherent states have recently become experimentally accessible \cite{Martinis}. It turns out that the case of a triplet of
coherent or squeezed states has the special property that its blind spots
form a regular pattern of displacements on an hexagonal lattice, whatever
the locations, amplitudes or phases of the three component states.
This structure is described in section 5.

In contrast to the robustness of the blind spots of pure states
with respect to any variation of parameters, they are remarkably sensitive
to decoherence. Indeed they will be shown in section 6
to be much more sensitive than the oscillations of the Wigner function,
whose disappearence is often taken as a sure indication
of loss of quantum coherence. One should note that the washing out of
fine oscillations of the Wigner function correspond to the decay of large
arguments of the characteristic function, whereas we here analyze
phenomena for small arguments. Thus, a density operator
may still have negative regions in its Wigner function,
even though any trace of its blind spots has been wiped out.

\section{Phase space representations}

Recall that, for systems with a single degree of freedom,
the corresponding classical phase space is a plane,
with coordinates $x = (p, q)$, momenta and positions.
The uniform translations of this space, $x\rightarrow x+\xi$,
define the {\it space of chords}, $\{\xi\}$.
The corresponding quantum translation (or displacement) operators, $\opT_\xi$,
transform position states, $|q\rangle\rightarrow|q+\xi_q\rangle$
(within an overall phase) and momentum states, $|p\rangle\rightarrow|p+\xi_p\rangle$.
For a general state, $|\Psi \rangle$, we have $|\Psi\rangle\rightarrow|\Psi_\xi\rangle$
and, in the special case of the ground state of the harmonic oscillator, this is transported into a coherent state:
$\opT_\xi\; |0\rangle =|\xi\rangle$.

Given the corresponding pure state density operators,
$\hat\rho = |\Psi \rangle\langle\Psi|$ and
$\hat\rho_\xi = |\Psi_{\xi} \rangle\langle\Psi_{\xi}|=\hat T_{\xi}\; \hat\rho \;\hat T_{-\xi}$,
the {\it quantum phase space correlations}, whose zeroes we seek,
are defined as the expected value for the translated state. In terms of the operator trace this is
\begin{equation}
C(\xi)= \tr\;\hat\rho\;\hat\rho_\xi= \tr\;\oprho\;\opT_\xi\;\oprho\;\opT_{-\xi}.
\end{equation}
The key point is that, for a pure quantum state,
$C(\xi)= |\langle\Psi|\Psi_{\xi}\rangle|^2$, so the blind spots
lie at intersections of nodal lines for both the real and the imaginary parts
of $\langle\Psi|\Psi_{\xi}\rangle$, which is generally a complex function of the chords.

In the present context, it is best to let the translation operators themselves
represent quantum operators.
In the case of density operators, $\chi(\xi)= \tr\; \opT_{-\xi}\; \oprho$ defines the (complex) chord function,
also known as the Weyl function, or as one of several quantum characteristic functions.
Note that the chord function has here been normalized so that $\chi(0) = \tr \hat \rho = 1$.
Sometimes used as a theoretical tool, it is a full representation of the density operator $\oprho$
and, in the case of a pure state, we have $\chi(\xi)= \langle\Psi|\Psi_{\xi}\rangle$,
i.e. the translation overlap itself \cite{ChouVou,LNP, Report}.
In terms of the position representation
\cite{ChouVou, AlmVal04}
\begin{equation}
\chi(\xi_q,\xi_p)=\int dq\; \langle q+\xi_q/2|\hat\rho|q-\xi_q/2\rangle \; e^{-i\xi_p\cdot q/\hbar}.
\end{equation}

A double Fourier transform then defines the celebrated Wigner function \cite{Wigner},
$W(x)={\rm FT}\{\chi(\xi)\}$, which is real, though it can have negative values.
The fact that the Wigner function can also be written as $W(x)= \tr\; \opR_x\; \oprho$,
reflects the Fourier relation of the translation operator itself to the reflection operator, $\opR_x$
\cite{Grossmann, Royer, Report}.
In terms of the position representation, we have
\begin{equation}
W(x) = W(q,p)= \frac{1}{2\pi\hbar}\int d\xi_q\; \langle q+\xi_q/2|\hat\rho|q-\xi_q/2\rangle \; e^{-i p\cdot \xi_q/\hbar}.
\end{equation}
The phase space correlation coincides with the correlation of the Wigner function
\cite{AlmVal04,LNP}, that is,
\begin{equation}
C(\xi)= 2\pi\hbar\int dx\; W(x)\; W(x-\xi) =
\frac1{2\pi\hbar}
\int {d\eta} \;e^{i\eta\wedge\xi/\hbar}\;|\chi(\eta)|^2,
\label{Wcor}
\end{equation}
just as if the Wigner function were an ordinary (positive) probability distribution
and $\chi(\xi)$ were its characteristic function. (Note that the two-dimensional vector product,
or {\it skew product} in the exponent; it is present in all double Fourier transforms in this article.)

The structure of zeroes of $C(\xi)$ is not obtained directly from those of the Wigner function.
Generically, the zeroes of $W(x)$ lie along nodal lines, because the Wigner function is real.
(In the case of more than one degree of freedom, this becomes a codimension-1 nodal surface
in the higher dimensional phase space.)
This is not the general case for the chord function:
No constraint obliges the nodal lines of the real and the imaginary parts
of $\chi(\xi)$ to coincide, so, typically, they intersect at isolated zeroes.
(In the general case, these become codimension-2 surfaces.)
We show below how the real and the imaginary parts of the chord function
are related respectively to the diagonal and the off-diagonal elements in a decomposition
of the density operator, with respect to the eigen-subspaces of the parity operator, $\opR_0$,
or any other reflection operator, $\opR_x$.

Reflection symmetry is not a precondition for the surprising property
of Fourier invariance of the phase space correlations for all pure states \cite{ChouVou,AlmVal04,LNP}:
${\rm FT}\{C(\xi)\}=C(\xi)$. This is simply a consequence of the fact that,
for pure states, we may combine $C(\xi)=|\chi(\eta)|^2$ with (\ref{Wcor}).
Thus a conjugacy between large and small scales
is generated, leading to tiny sub-Planck structures \cite{Zurek}
for quantum pure states with sizeable correlations over an area
that is appreciably larger than Planck's constant, $\hbar$.
In short, large structures imply large wave vectors for the oscillations
of $W(x)$ and for the oscillations of the real and the imaginary parts of $\chi(\xi)$.
These real functions have nodal lines which must approximate the origin,
as the overall extent of the state is increased.
It remains to show that generally these nodal lines intersect transversally.

These considerations are easily adapted to systems with several degrees of freedom:
The origin of the higher dimensional phase space lies in the codimension-1 nodal surface
of the imaginary part of the chord function
and this generically intersects the nodal surfaces of the real part of the chord function
along codimension-2 surfaces. Therefore the overlap zeroes for translated copies of the state
will generically give rise to isolated blind spots in arbitrary two-dimensional sections of
the higher dimensional phase space.
Thus, the likelyhood that a continuous group of translations,
evolving from the origin along a straight line in the space of chords,
comes close to a point of zero overlap does not depend on the number of degrees of freedom
and it is certain to occur for centro-symmetric systems.
Generically, blind spots lie close to the origin for all extended systems,
because the invariance of the phase space correlations with respect to Fourier transformation
holds, regardless of the number of degrees of freedom.

\section{Real and imaginary nodal lines}

The reflection operators, $\hat R_x$, may be considered to be `generalizations' of the parity operator,
$\hat R_0$, which corresponds classically to the phase space map: $x\to-x$.
Because $\hat R_x$ is Hermitian (as well as unitary) and $(\hat R_x)^2=1$, there are only two eigenvalues,
$\pm 1$, which define a pair of orthogonal even and odd subspaces of the Hilbert space, for each choice of
the reflection centre, $x$. Hence,
any pure state $|\psi\rangle$ may be decomposed into a pair of parity-defined components,
an odd $|\psi_o\rangle$ and an even $|\psi_e\rangle$,
i.e. $|\psi_e\rangle$ and $|\psi_o\rangle$ are eigenstates of $\hat R_0$,
for $+1$ and $-1$, respectively. The corresponding pure state density operator
can then be decomposed into a pair of components, $\hat\rho_D$ and $\hat\rho_N$, defined by:
\begin{equation}
\hat\rho=\hat\rho_D+\hat\rho_N=(|\psi_o\rangle\langle\psi_o|+|\psi_e\rangle\langle\psi_e|)+(|\psi_o\rangle\langle\psi_e|+|\psi_e\rangle\langle\psi_o|).
\label{rho-parity}
\end{equation}
This decomposition is generalized to mixed states by just adding the diagonal,
or nondiagonal contributions from each pure state in the mixture.

It is well known that the value of the Wigner function at each point, $x$,
is defined in terms of the diagonal component \cite{Royer, LNP}:
\begin{equation}
W(x)= \frac{1}{\pi \hbar}[\langle \psi_e|\psi_e\rangle - \langle \psi_o|\psi_o\rangle].
\end{equation}
Furthermore, for $x=0$, these `diagonal' terms making up $\hat\rho_D$ have a definite parity,
i.e. $[\hat R_{ 0},|\psi_o\rangle\langle\psi_o|]=[\hat R_{ 0},|\psi_e\rangle\langle\psi_e|]=0$ .
Since the chord representation of such a centro-symmetric density operator is a mere rescaling
of its Wigner function, it follows that the diagonal component of the chord function is entirely real.

\begin{figure}[htb!]\centering\includegraphics[width=15.7cm]{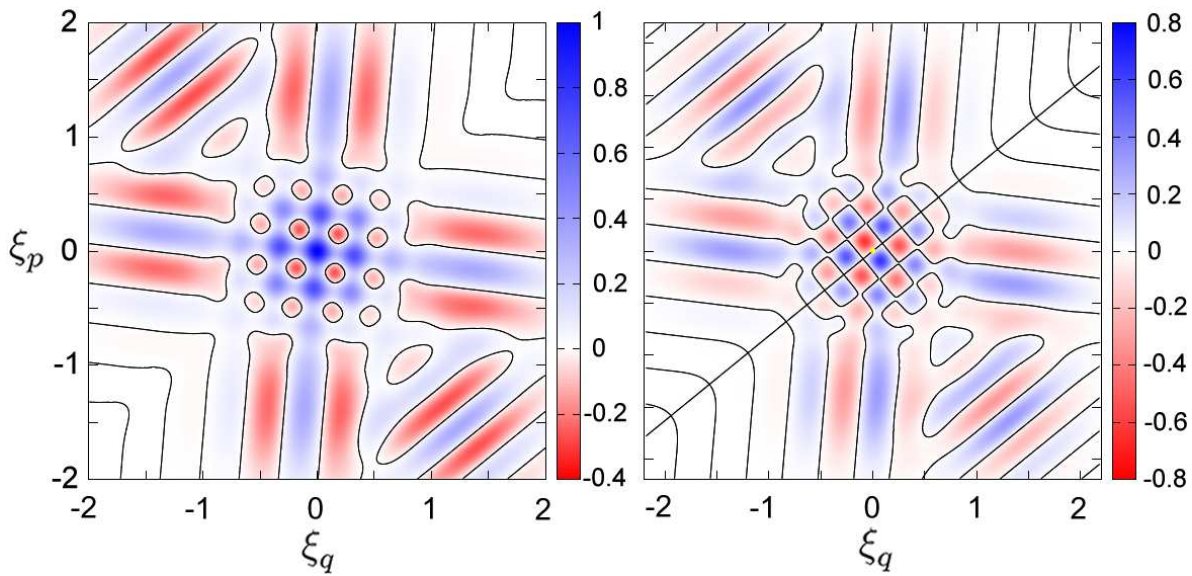}\caption{\textbf{The real (\emph{left}) and imaginary (\emph{right}) parts of the chord function for a coherent state triplet}. Note  the definite even (odd) parity of the real (imaginary) part. The blind spots are located at the intersection of both two sets of nodal lines (continuous black lines).} \label{f4}\end{figure}

This does not necessarily imply that the chord representation of $\hat\rho_N$ is purely imaginary,
but this is indeed the case. To see this, recall that \cite{Report} $\hat R_{ 0}\hat T_\xi\hat R_{ 0}=\hat T_{-\xi}$.
Hence, denoting  $\chi_{N}(\xi)=\tr\hat T_{-\xi}(|\psi_o\rangle\langle\psi_e|+|\psi_e\rangle\langle\psi_o|)$,
we obtain
\begin{equation}
\fl \chi_{N}(\xi)=\langle\psi_o|\hat T_{-\xi}|\psi_e\rangle+\langle\psi_e|\hat T_{-\xi}|\psi_o\rangle=
\langle\psi_o|\hat R_{ 0}\hat T_{\xi}\hat R_{ 0}|\psi_e\rangle+
\langle\psi_e|\hat R_{0}\hat T_{\xi}\hat R_{0}|\psi_o\rangle=
-{\chi^*_{N}(\xi)},
\end{equation}
where asterisk indicates the complex conjugate.
Since $\chi(-\xi)=\chi^*(\xi)$, it follows that the real and the imaginary parts of the chord function are even and odd functions, respectively.
Unlike the value of the Wigner function, 
that depends exclusively on the
diagonal component, $\hat\rho_D$, the chord function depends also on the nondiagonal component, $\hat\rho_N$.
Indeed, the latter determines exclusively the imaginary part of the chord function,
whereas $\hat\rho_D$ accounts entirely for the real part. These are shown in figure \ref{f4}
for the superposition of three coherent states.

The imaginary part is zero at the origin,
because $\tr \hat\rho_N=0$, so it always lies on an imaginary nodal line.
In contrast, the real part of the chord function attains there its maximum value, for $\tr \hat\rho_D=1$.
The generic scenario is then that an imaginary nodal line across the origin intersects transversally
the real nodal lines, which avoid the origin. 

For a centro-symmetric state, the imaginary part is identically zero, leading to continuous lines of zero overlap, but any slight
symmetry breaking isolates the zeroes into blind spots. Nonetheless, for nearly centro-symmetric
states, $|\chi(\xi)|^2$ will remain very small along its real nodal lines.
The case of the Schr\"odinger cat state is specially pathological:
Breaking the symmetry of the coefficients of the pair of coherent states,
generates straight nodal lines of the imaginary part of the chord function
that are parallel to the real nodal lines. Hence, the blind spots vanish entirely.

It may be pondered that the parity decomposition into even and odd subspaces
is entirely dependent on the choice of origin as a reflection centre. Therefore, the
real and imaginary parts of the chord function are not invariant with respect to a shift of origin.
Indeed, the effect of such a translation by $\eta$ is merely the multiplication of a phase factor:
\begin{equation}
\chi_\eta(\xi)=\exp\left(i\frac{\eta\wedge\xi}\hbar\right) \chi(\xi).
\end{equation}
Nonetheless, this phase factor does not alter the blind spots, because we can still
decompose the chord function into the same pair of (phase shifted) components and
their nodal lines still intersect at the same blind spots, resulting from the intersection of the nodal lines in figure \ref{f4}.

Centro-symmetric states, i.e. those for which the commutator $[\oprho, \opR_x]=0$,
for some reflection centre, $x$, are thus the important exceptions to nodal lines that intersect transversely.
In these cases, the chord function is essentially a real rescaling of the Wigner function \cite{LNP},
so that its zeroes also lie along nodal lines, rather than isolated points.

\section{Superpositions of coherent states}

General coherent states allow for further quantum unitary transformations,
corresponding to linear classical transformations, beyond the translations which define them.
Leaving this possibility as implicit, we merely denote the superposition as
\begin{equation}
|\Psi^N \rangle= a_0 |0\rangle+ a_1 |\eta_1\rangle+...+a_N |\eta_N\rangle.
\end{equation}

Superconducting resonators and nonlinear optical processes, among others techniques \cite{Martinis,tec1,tec2,tec3} can generate these kind of states.
A translation of this state is then
\begin{equation}
|\Psi_{\xi}^N \rangle= a_0 |\xi\rangle+ a_1 e^{i\xi\wedge\eta_1/\hbar} |\eta_1+\xi\rangle+...+a_N e^{i\xi\wedge\eta_N/\hbar} |\eta_N+\xi\rangle,
\end{equation}
within an overall phase.
So, in the correlation, $C(\xi)$, we deal with $(N+1)^2$ terms of the form $\langle\eta_n|\;\eta_m +\xi \rangle$.
One should note that each diagonal term, $\langle\eta_n|\;\eta_n +\xi \rangle$,
determines the internal correlations for the single coherent state $|\eta_n\rangle$.

The wave function for coherent states
of the quantum harmonic oscillator of frequency $\omega$ is
\begin{equation}
\psi(q)\equiv\langle q|\eta\rangle=
\left(\frac{\omega}{\pi\hbar}\right)^{1/4}
\exp\left[-\frac{\omega}{2\hbar}(q-\eta_q)^2+\frac{i}{\hbar}\eta_p\left(q-\frac{\eta_q}2\right)\right],
\end{equation}
so that the chord function for a coherent state is
\begin{equation}
\hspace{-2cm}
\chi_\eta(\xi)=\exp\left(i\frac{\eta\wedge\xi}\hbar\right)
\exp\left[-\frac{\omega}{\hbar}\left(\frac{\xi_q}{2}\right)^2
          -\frac1{\hbar\omega}\left(\frac{\xi_p}{2}\right)^2\right]
\stackrel{\omega=1}{\longrightarrow}e^{i\eta\wedge\xi/\hbar}e^{-\xi^2/4\hbar},
\end{equation}
whereas its Wigner function is
\begin{equation}
W_\eta(q,p)=\frac{1}{\pi\hbar}
\exp\left[-\frac{\omega}{\hbar}(q-\eta_q)^2-\frac1{\hbar\omega}(p-\eta_p)^2\right]
\stackrel{\omega=1}{\longrightarrow}\frac{1}{\pi\hbar}e^{-(x-\eta)^2/\hbar}.
\end{equation}
Both the Wigner function and the chord function are transformed classically
by linear canonical transformations,
so generalized coherent states are also two-dimensional Gaussians and the elliptic level curves for $W_\eta(x)$ and for $|\chi_\eta(\xi)|^2$
can be rotated as well as elongated.

In the case of a multiplet of coherent states, both the Wigner function and the chord function
split up into diagonal and nondiagonal terms: The former, $W_{n}(x)$, are well known to have Gaussian
peaks at each coherent state centre $\eta_n$, while $W_{nm}(x)$ is represented by interference fringes,
localized halfway between $\eta_n$ and $\eta_m$.
This pattern is structurally stable with respect to variations of the coefficients of $|\Psi^N \rangle$. Note that this decomposition for the Wigner and the chord function gives the false impression that individual pairs of coherent states are the fundamental building blocks for their full phase space picture and so for the blind spots. However, we shall show that the distribution of the blind spots changes drastically by adding a single coherent state to the initial multiplet.

By focusing on very short ranges, one can dissect the simple skeleton
that supports the pattern of blind spots. First, consider the diagonal terms
of the full Wigner function, i.e. the Gaussian peaks $W_{n}(x)$.
Even if they are squeezed in different ways, all their widths
will be $\GO(\hbar^{1/2})$, where we consider $\hbar$ as a small parameter,
that is, the widths are much smaller than the separations,
$|\eta_n-\eta_m|$, which are $\GO(\hbar^{0})$.
Altogether, this positive part of the full $W^N(x)$ can be identified with the Wigner function
for the {\it mixed quantum state} obtained from the same set of generalized coherent states:
$|a_0|^2 W_0(x)+ |a_1|^2 W_1(x-\eta_1)+...+|a_N|^2 W_N(x-\eta_N)$.
The corresponding mixed state chord function, resulting from a Fourier transform, is then
\begin{equation}
\sum_{n=0}^N \;|a_n|^2\; \chi_n(x)\; \exp\left(\frac{i}{\hbar}\eta_n \wedge \xi \right).
\label{chordmix}
\end{equation}
Just as $W_n(x)$, each of the individual coherent state chord functions, $\chi_n(\xi)$,
has a width $\GO(\hbar^{1/2})$, though it is centred on the origin.
Therefore, on a tiny scale $\GO(\hbar)$, we may approximate
these by infinite widths, leading to the approximate chord function:
\begin{equation}
\chi^{\delta}(\xi)= \sum_{n=0}^N \;|a_n|^2\; \exp\left(\frac{i}{\hbar}\eta_n \wedge \xi \right),
\label{dif}
\end{equation}
which is independent of the relative phases in the superposition $|\Psi^N\rangle$.
This expression is analogous to the amplitude of the (far field) diffraction pattern
from point scatterers with weights $|a_n|^2$, placed at $\eta_n$ (except for a $\pi/2$ rotation,
because of the vector product). Squaring for the {\it diffraction intensities},
we obtain the {\it small chord approximation}, $C(\xi)\rightarrow C^\delta(\xi)=|\chi^\delta(\xi)|^2$,
so that the approximate zeroes can be read off directly from $\chi^\delta(\xi)$. We remark that a moment expression, such as adopted in \cite{Alonso}, cannot be extended to the  calculation of blind spots, such as obtained here.

\begin{figure}[htb!]\centering\includegraphics[width=15.7cm]{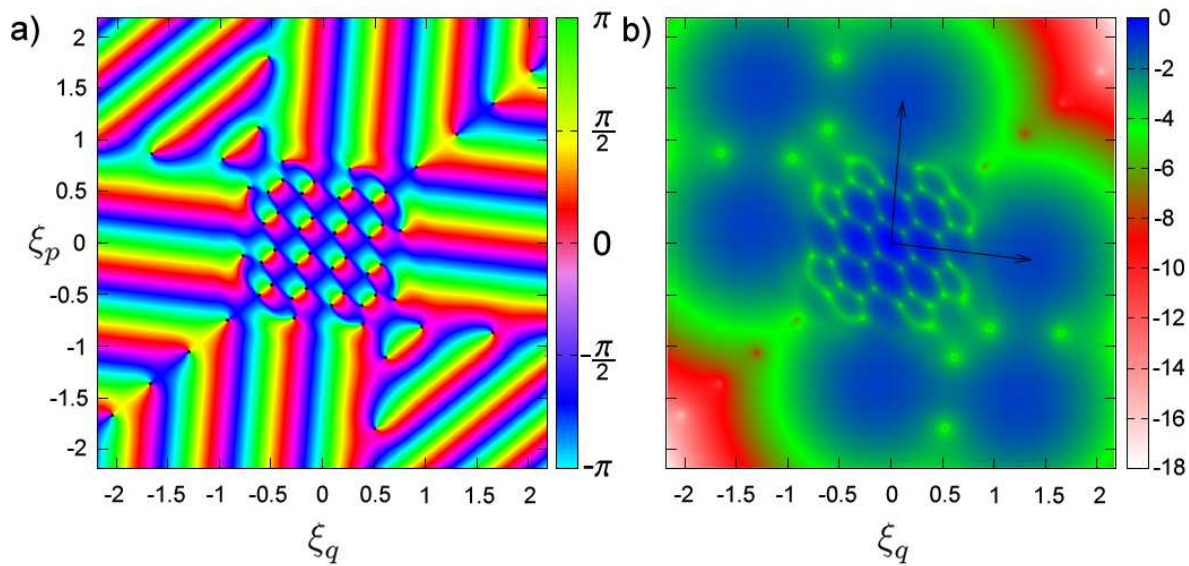}\caption{\label{f1}
\textbf{Logarithmic plot of the chord function for a coherent state triplet}: phase a) and intensity b). Large scale correlations show up as a hexagon of (complex) Gaussians (big blue disks in b)), while the diagonal terms produce the pattern around of the origin. The black points (phase singularities) in a), or the maxima in b), are the blind spots, forming a `honeycomb' lattice near the origin. The arrows locate the states forming the triplet.}\end{figure}

Note that each zero corresponds to a singularity of the phase of the chord function
and, hence, a wave dislocation \cite{NyeBerry}. The overall phase pattern
for the chord function of a coherent state triplet ($\hbar=0.075$)
with equal weights and vertices located at $(0,0),$ $(1.5,-0.1)$ and $(0.2,1.5)$ is shown in figure \ref{f1}a.
The singularities at the blind spots are shown as black points in the figure \ref{f1}a:
Their neighbourhood contains all possible phases.
The corresponding logarithmic plot of the intensities is presented in figure \ref{f1}b.

There would be nothing to wonder at, if \eref{dif} really represents a diffraction pattern.
It is the alternative interpretation as a field of correlations, for the same source,
that assumes a paradoxical flavour:
The zeroes of this simple pattern signal the neighbouring presence of blind spots in
the overlap of the original multiplet of coherent states with its translations.
The relevance of the `diffraction pattern' for the correlation
lies in the latter's Fourier invariance.
For a true mixture of coherent states, the correlations would be exclusively the Fourier transform
of the chord intensity (not equal to $|\chi^\delta(\xi)|^2$),
with the result that the zeroes would be washed away.
It is the outer part of the full pure state chord function, shown in figure 1,
which was neglected, that guarantees the Fourier invariance for $C(\xi)$,
notwithstanding its negligible direct effect on the diffraction pattern itself.
Indeed, each nondiagonal term, $\chi_{nm}(\xi)$, contributing to the chord function
is a (complex) Gaussian, also of width $\GO(\hbar^{1/2})$, centred on $\eta_n-\eta_m$.
Together they form the outer hexagon in figure \ref{f1}b, corresponding to the simple external oscillations in figure \ref{f1}a.
The condition that the separation of all pairs of coherent states is $\GO(\hbar^{0})$,
implies that these outer maxima account for only an exponentially small perturbation
of the central diffraction pattern.

\section{Hexagonal lattice of blind spots for a coherent state triplet}

We have seen that the commonly chosen case of $N=1$ is nongeneric,
i.e. Schr\"odinger cat states have either a continuum of blind spots, or none at all.
Generically, the correlation zeroes are isolated,
but they are only disposed in a periodic pattern for $N=2$.
To see this, consider each term in equation (\ref{dif}) as a vector of length $|a_n|^2$ in the complex plane.
Then the condition for a given chord, $\xi_0$, to specify a blind spot
is that the $N+1$ vectors form a closed polygon.
However, a triangle is the only polygon that is completely determined
(within discrete symmetries) by the lengths of its sides.
For large $N$, the statistical properties
of the distribution of zeroes will be approximately those of random waves, but these are complex,
so that the blind spots are the intersections of real and imaginary random nodal lines \cite{Longuet}.

\begin{figure}[htb!]\centering\includegraphics[width=15.7cm]{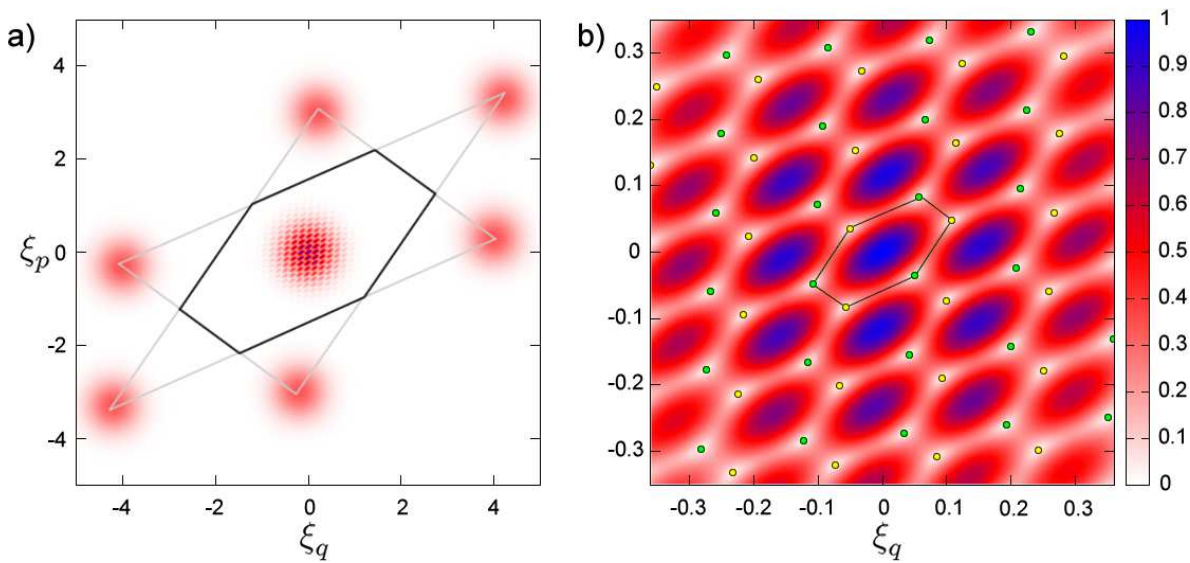}\caption{\textbf{Intercalated quasi-orthogonal sublattices}. a) Linear intensity of the chord function and b) its magnification in the neighbourhood of the origin.The hexagonal lattice of blind spots is formed by the overlap of two oblique sublattices (green and yellow spots respectively). The black hexagon in a) is a rescaling of the hexagons of the blind spots lattice.Here the states are located at $(0,0),$ $(-4,0.3)$ and $(0.2,3)$, $\hbar=0.075$ and the amplitudes in the superposition are uniform.\label{f2}}\end{figure}

For $N=2$, a triplet of coherent states, the three intensities, $|a_n|^2$,  determine an unique triangle in the complex plane. Since the first vector has been chosen with the direction $\theta_0=0$, the other angles are given by the law of cosines, i.e.
\begin{equation}
\theta_1=\pi\pm\textrm{ arccos} \left[\frac{|a_0|^2+|a_1|^2-|a_2|^2}{2 |a_0a_1|}\right],
\label{theta}
\end{equation}
for allowed values of arccos (and likewise $\theta_2$ by an index permutation).
For each choice of sign in equation \eref{theta}, the solution of the pair of linear equations,
\begin{equation}
\theta_n+ 2\pi k_n = \frac{\eta_n \wedge \xi_0}{\hbar}\; ,
\label{linear}
\end{equation}
determines two oblique sublattices of blind spots, $\xi_0(k_1,k_2)$: The green and the yellow points in figure \ref{f2}b.
Together these form the full hexagonal lattice, shown in figure \ref{f2}.
Since the vectors between the sublattices equal the vectors from the chord origin to either sublattice,
any superposition of triplets located at the ``green spots" in the figure \ref{f2}b is quasi-orthogonal to
superpositions of triplets located at ``yellow spots".

The {\it star of David} inscribed in the hexagon of large scale peaks in figure \ref{f2}a
defines, in its turn, an interior hexagon, which is just a rescaling
of the blind spot hexagons of the lattice in figure \ref{f2}b.
This a prime example of the conjugacy of large and small scales generated
by Fourier invariance of the correlations.
For well separated coherent states, such that components of $\eta_n-\eta_m$ are $\GO(\hbar^{0})$,
the deviations of $C(\xi)$ from the lattice are minute, as shown in figure \ref{f2}b,
that is, each small green or yellow `spot' contains a true zero.
The small deviations can be easily calculated by Newton's method.

It is important to point out that the coefficients of a superposition of states
remains invariant during an unitary evolution, while, for instance, the positions
of generalized coherent states are moving.
Thus, an experimental measurement of a pair of correlation zeroes, $\xi_0(t)$,
at any time, can in principle be fed into the pair of equations (\ref{linear})
to solve the inverse problem, that is, to obtain the coherent state positions: $\eta_n(t)$.

\section{Decoherence}

Interaction with an uncontrolled external environment evolves a pure state
into a mixed state. The Wigner function of an extended state gradually looses
its interference fringes and becomes everywhere positive in all cases \cite{BroAlm04,DioKief},
be it a multiplet of coherent states, or an excited eigenstate of an anharmonic oscillator.
Viewed in the chord representation, this effect is translated into the loss of amplitude
of $\chi(\xi)$ for all chords outside a neighbourhood of the origin of area $\hbar$.
However, we are now focussing on structures that lie within this supposedly {\it classical region}.
So, even though blind spots and the oscillations that give rise to them
are a delicate quantum effect, it is not a priory clear how they are affected by decoherence.

The crucial point is that
it is no longer possible to identify the square of the chord function
with the phase space correlations, that is, we must use $C(\xi)={\rm FT}\{|\chi(\xi)|^2\}$ for mixed states.
Even though, the theory of open quantum systems is in no way as complete as the quantum description
of unitary evolution, appropriate for closed systems, it will be here established that the observable correlations,
$C(\xi)$, are exceptionally sensitive to decoherence, even though $\chi(\xi)$ preserves
its oscillatory structure and its multiple zeroes.

The full evolution of any quantum system coupled to a larger system, in the role of environment,
is still unitary and it can be resolved into the completely positive evolution of the reduced system,
by tracing away the environment \cite{Giulini}.
In the limit of weak coupling, the reduced system evolution becomes memory independent
({\it Markovian}) and is governed by the {\it Lindblad master equation} \cite{Lindblad},
\begin{equation}
\frac{\partial\hat\rho}{\partial t}=
\frac{i}\hbar[\hat H,\hat \rho]-\frac{1}{2\hbar}\sum_j\left(2\hat L_j\hat\rho\hat L^\dag_j
-\hat L^\dag_j\hat L_j\hat\rho-\hat\rho\hat L^\dag_j\hat L_j\right),
\end{equation}
where $\hat L_j$ is a {\it Lindblad operator}, which models the interaction between the system and the environment.

Assuming that each operator $\hat L_j$ is a linear function of the momentum $\hat p$ and position $\hat q$ operators
and that the Hamiltonian, $\hat H$, is quadratic, leads to the chord phase space equation \cite{BroAlm04},
\begin{equation}
\fl \frac{\partial \chi_t(\xi)}{\partial t}=
\{H(\xi),\chi_t(\xi)\}-\alpha\;\xi\cdot\frac{\partial \chi_t(\xi)}{\partial t}
-\frac{1}{2\hbar}\sum_j[(l_j'\cdot\xi)^2+(l_j''\cdot\xi)^2] \;\chi_t(\xi).
\end{equation}
Here $\{\cdot,\cdot\}$ is the classical Poisson bracket, $L_j(x)=(l_j'+ i l_j'')\cdot x$ and $H(x)= x\cdot \mH \cdot x$
are the Weyl representation of $\hat L_j$ and $\hat H$ and $\alpha\equiv\sum_jl_j''\wedge l_j'$ is the {\it dissipation coefficient}. The solution for this equation is given by a product of two factors.
One is the unitarily evolved chord function,
$\chi_u(\xi, t)= \chi(\xi(t))$, which propagates classically as a Liouville distribution,
just as the Wigner function. This is multiplied by a Gaussian function, $\exp(-\xi\cdot \M_t\cdot\xi/\hbar)$,
that depends only on the Hamiltonian and the Lindblad coefficients, but not on the initial chord function.
The explicit general form,
\begin{equation}
\M_t=2\sum_j\int_0^tdt'\;e^{2\alpha(t'-t)}\;\R_{t'-t}^T l_j l_j^T \R_{t'-t},
\end{equation}
where
\begin{equation}
\R_t=\exp(2 \J \mH t)
\;\;\;\;\textrm{and }\;\;\;\;
\J=\left(
\begin{array}{cc}
    0 & -1 \\
    1 &  0
\end{array}
\right),
\end{equation}
is given in \cite{BroAlm04}, but various special cases have been previously obtained
\cite{Agarwal, DodMan, SanScuScheid, DioKief}.

The possibility of deducing the complete local structure of the correlations, $C(\xi,t)$,
directly from that of the chord function still exists
for this special but important class of quantum Markovian evolutions.
The reason is that $|\chi(\xi, t)|^2$ also factors, such that $|\chi_u(\xi, t)|^2$
is multiplied by the squared Gaussian. A Fourier transform supplies the evolved correlation as merely the convolution,
\begin{equation}
C(\xi, t)=|\chi_u(\xi,t)|^2 *\; \exp \left[-\frac{1}{4\hbar}\xi \cdot {\M_t}^{-1}\cdot \xi \right],
\end{equation}
because $\chi_u(\xi)$ still represents a pure state and,
hence, ${\rm FT}\{|\chi_u(\xi,t)|^2\}= |\chi_u(\xi,t)|^2$.
\begin{figure}[htb!]\centering\includegraphics[width=15.7cm]{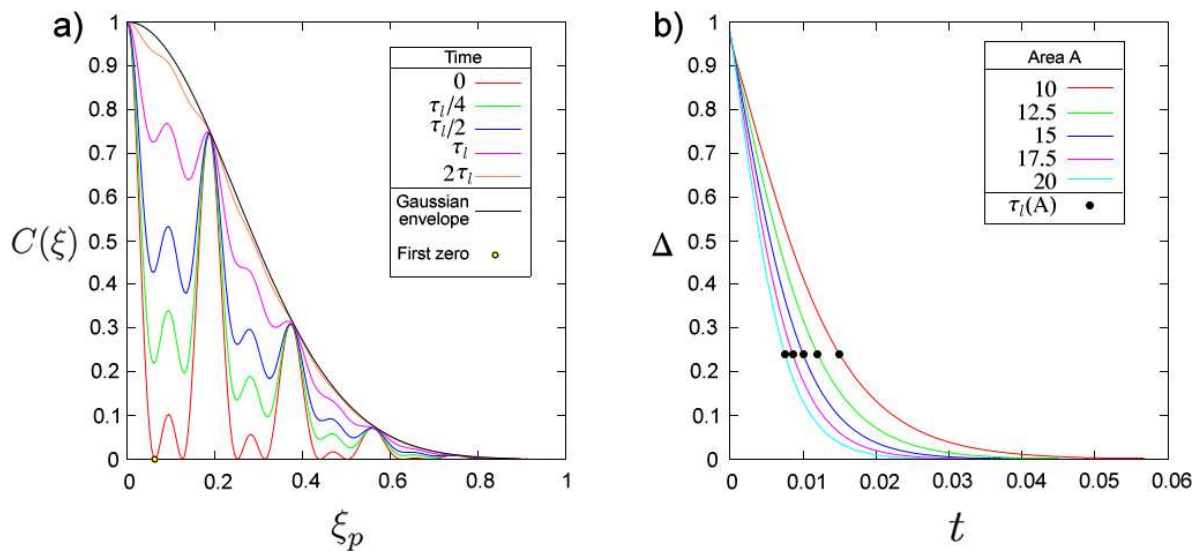}\caption{\textbf{Effect of decoherence on correlation zeroes}. Markovian evolution of the phase space correlation for a triplet located at $(0,0)$, $(0,d)$ and $(d,0)$, along the line $\xi_p=-\xi_q/2$. In this case $\chi_u(\xi,t)=\chi(\xi,0)$ and the Lindblad operators are $\hat p$ and $\hat q$. In a) the zeros are seen to be lifted with increasing time and even the local minima are distinguishable only until the lifting time $\tau_l$. Here, $d=5$, so the area of the triplet is $A=12.5$ and $\hbar=0.075$. b) shows the difference, $\Delta$, between the phase space correlation and its Gaussian envelope as a function of time, at the first zero, for the same configuration of a), but varying the area, $A=d^2/2$, of the triplet.} \label{f3}\end{figure}

As a result of this coarsegraining of the positive function, $|\chi_u(\xi, t)|^2$,
the correlation zeroes are lifted in time, as illustrated in figure \ref{f3}.
Nonetheless, they can still be detected as local minima,
remaining up to a {\it lifting time}, $\tau_l$, which depends on the disposition of the coherent states,
as shown in figure \ref{f3}b.
This time is reached when the correlation, $C(\xi,t)$,  approaches its Gaussian envelope.
Simple dimensional considerations lead to the definition
\begin{equation}
\tau_l=\frac{\hbar}{A} t_p,
\label{tauratio}
\end{equation}
where $A$ is the area of the triplet and $t_p$ is the {\it positivity time} of the Wigner function
\cite{DioKief, BroAlm04}. One should note that, though $t_p$ depends on the Lindblad coefficients,
the ratio $\tau_l/t_p$ is hardly affected by them.

We have here chosen a very special example where $\chi_u(\xi, t)$ does not evolve, so as to keep
the correlation minima over the blind spots. Nonetheless, the relation between the different decoherence times
(\ref{tauratio}) holds much more generally, even if the area, $A(t)$, for the evolving extended state
is not constant. Recalling that $t_p$ measures the survival time of the interference fringes in the Wigner function,
we find that the survival time for the correlation minima that would indicate `nearly' blind spots for open systems,
is much smaller if $A\gg\hbar$. Thus, the oscillatory structure that gives rise to blind spots for small
displacements is much more sensitive to decoherence than the fringes of the Wigner function.

\section{Discussion and outlook}

Quasi-orthogonality between a state and its slight evolution
is a remarkable quantum property with
applications in the theory of decoherence \cite{Alonso} and quantum metrology \cite{ToscZur}.
However, the methods in previous studies did not settle the question of whether
complete orthogonality is also accessible. We have shown how the interplay between the chord function,
reflection symetries and the remarkable Fourier invariance of the phase space correlations
determine general srtuctures that are valid for any particular example.
We have here given special attention to coherent state triplets.
They lie in the generic class where the blind spots are isolated,
but their ordering in the neighbourhood of an ordered lattice is unique.

The decay of overlap for a translation is the simplest instance of the
general loss of fidelity for a pair of quantum states, $|\Psi(t) \rangle$ and $|\Psi'(t) \rangle$,
that evolve under the action of slightly different Hamiltonians.
Most treatments of this process have been based on the unitary evolution
of initial single coherent states \cite{JalPas, Gorinetal}. We have here supplied an example, where
the addition of further structure to the initial state modulates the average overlap decay
in an unanticipatedly rich manner. For a short time, we may even approximate
such a more general evolution of a multiplet, through a local quadratic semiclassical theory \cite{HellerLH},
to continue to be a superposition of generalized coherent states.
Generally, the overlap $\langle\Psi(t)|\Psi'(t)\rangle$
will be a complex function for all time, so that zeroes (for a pair of parameters)
will be isolated and they can only be narrowly avoided as the pair of states evolve in time.

Some decoherence is needed for the fidelity of structured states to decay smoothly.
The indication of this from Markovian systems with quadratic Hamiltonians and linear coupling operators
can be somewhat generalized in a WKB-type approximation, which still produces evolution
of the correlations in the form of a convolution, though the coarsegraining window
is no longer Gaussian \cite{OzRiBro}. In any case, the present example has brought to light
the contrast between a remarkable robustness of blind spots
as regards unitary evolution and their extreme sensitivity to decoherence.

\ack

We thank Ra\'ul Vallejos, Fabricio Toscano and Marcos Saraceno for
useful discussions; the latter and Edgardo Brigatti for valuable
computational advice. Partial financial support from INCT-IQ, CNPq
and CAPES-COFECUB is gratefully acknowledged.
\newline

\end{document}